\documentclass[12pt]{article}
\usepackage{graphicx}
\usepackage{subfigure}
\usepackage{amssymb}
\usepackage{amsfonts}
\usepackage{latexsym}

\usepackage[usenames]{color}
\usepackage{latexsym}
\usepackage{epstopdf} 
\usepackage[normalem]{ulem}

\newcommand{\beq}{\begin{equation}}
\newcommand{\eeq}{\end{equation}}
\newcommand{\be}{\begin{equation}}
\newcommand{\ee}{\end{equation}}
\newcommand{\bea}{\begin{eqnarray}}
\newcommand{\eea}{\end{eqnarray}}

\makeatletter

\@addtoreset{equation}{section}
\makeatother

\def\href#1#2{#2}

\usepackage{amsmath}	
\begin{document}

\begin{titlepage}
\begin{flushleft}
       \hfill                       FIT HE - 21
-01 \\
       \hfill                       
\end{flushleft}

\begin{center}
  {\huge Stiff equation of state for a holographic \\
   \vspace*{2mm}
nuclear matter as instanton gas 
\vspace*{2mm}
}
\end{center}

\begin{center}

\vspace*{5mm}
{\large ${}^{\dagger}$Kazuo Ghoroku\footnote[1]{\tt gouroku@fit.ac.jp},
${}^{\dagger}$Kouji Kashiwa\footnote[2]{\tt kashiwa@fit.ac.jp},
Yoshimasa Nakano\footnote[3]{\tt ynakano@kyudai.jp},\\
${}^{\S}$Motoi Tachibana\footnote[4]{\tt motoi@cc.saga-u.ac.jp}
and ${}^{\ddagger}$Fumihiko Toyoda\footnote[5]{\tt f1toyoda@jcom.home.ne.jp}
}\\

\vspace*{2mm}
{${}^{\dagger}$Fukuoka Institute of Technology, Wajiro, 
Fukuoka 811-0295, Japan\\}
\vspace*{2mm}
{${}^{\S}$Department of Physics, Saga University, Saga 840-8502, Japan\\}
\vspace*{2mm}
{${}^{\ddagger}$Faculty of Humanity-Oriented Science and
Engineering, Kinki University,\\ Iizuka 820-8555, Japan}
\vspace*{3mm}
\end{center}

\begin{center}
{\large Abstract}
\end{center}

In a holographic model, which was used to investigate the color superconducting phase of QCD,
a dilute gas of instantons is introduced to study the nuclear matter.
The free energy of the nuclear matter is
computed as a function of the baryon chemical potential in the probe approximation.
Then the equation of state  is 
obtained at low temperature. Using the equation of state for the nuclear matter, 
the Tolman-Oppenheimer-Volkov equations for a cold compact star are solved.
We find the mass-radius relation of the star, which is 
similar to the one for quark star. This similarity implies that the instanton gas given here is a kind of self-bound matter.

\noindent

\begin{flushleft}

\end{flushleft}
\end{titlepage}
\newpage
\newpage

\section{Introduction}

\begin{figure}[htbp]
\vspace{.3cm}
\begin{center}
\includegraphics[width=10.0cm,height=7cm]{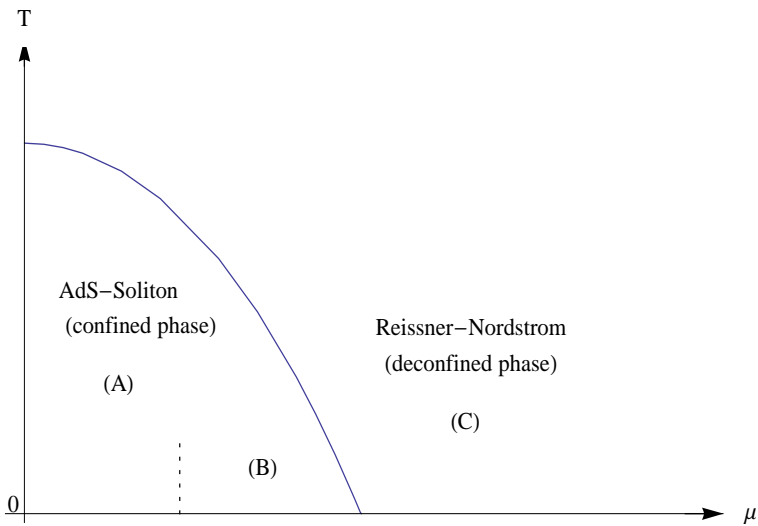}
\caption{A schematic phase diagram of QCD in ($\mu, T$) plane.
\label{Phase-diagram}}
\end{center}
\end{figure}

Many analyses of the holographic quantum chromodynamics (QCD) imply a schematic phase diagram shown by 
Fig.\,\ref{Phase-diagram}, 
where $\mu$ and $T$ denote the chemical potential of quark
and the temperature of the system, respectively. 
The solid curve represents
the confinement/deconfinement transition, 
which is obtained as the Hawking-Page transition points
from the anti-de Sitter (AdS)-soliton to the Reissner-Nordstrom (RN)
background configurations through the following bulk action 
\beq\label{B}
S_\mathrm{Bulk} = \int d^{d+1} x \sqrt{-g} \left\{ {\cal R} + {d(d-1) \over L^2}- {1 \over 4} F^2 \right\}\, 
\eeq
for $d=5$ \cite{Basu, GKNTT}.
This bulk action is written by the Einstein-Hilbert action with a negative cosmological constant, $d(d-1)/L^2$,
 and $(d+1)$-dimensional $U(1)$ gauge field, $F^2\equiv F_{\mu\nu}F^{\mu\nu}$.
The first gravitational part is proposed as a model dual to the Yang-Mills
(YM) theory with strongly interacting flavor fermions,
and the gauge field part
is dual to the baryon number current. 

We notice that, for the bulk background 
of $d=5$, 
the dimension of the boundary spacetime is effectively 3+1
since one space dimension is compactified by the Sherk-Schwarz compactification.
The above action has been firstly used to examine the electric superconductivity for $d=3$
\cite{Gub, Hart, Hart2, Nishi, Iqbal:2010eh} and theory with $R$-symmetry
for any $d$ \cite{Cham}. 

Recently, Eq.(\ref{B}) has been used as a bottom-up holographic QCD model to study
color superconductivity  \cite{Basu, GKNTT, Fadafan, Nam} 
in the region (C) of Fig.\ref{Phase-diagram}.
The analysis has been performed under the assumption that the $U(1)$ gauge field is dual to the baryon number current. The model is 
extended up to small $N_c$ where $N_c$ is the number of colors and the authors of Ref.\,\cite{GKNTT}
pointed out that the color superconducting (CSC) phase could not be found for $N_c\geq 2$. 
However, the existence
of the CSC phase has been 
pointed out when the model is improved by adding the higher derivative terms, the Gauss-Bonnet term, to Eq.(\ref{B}) \cite{Nam}. 

Another point to be noticed  is the  
dotted line shown in Fig.\,\ref{Phase-diagram}.
In the confined phase (A), baryons are constructed from $N_c$ quarks.
Then, at a certain chemical potential, they are assembled as nuclear matter.
The region (B) is called as the nuclear matter phase. 
In the previous studies, this phase has been foreseen 
in terms of the top-down model \cite{Bergman, RSRW, GKTTT}, based on the
Sakai-Sugimoto D4/D8 model \cite{SS,HSSY}. In these models, baryon is introduced as an instanton which is 
a soliton constructed from the flavored vector mesons.

Here, based on our bottom-up holographic model (\ref{B}), we study the nuclear matter 
according to the idea 
that it is constructed from the instanton gas.
In order to provide baryons, we
add $SU(N_f)$ gauge field sector to Eq.(\ref{B}) 
\cite{Bergman,SS,HSSY,Schmitt}, where $N_f$ is the number of flavors.
Then 
the dilute gas of instantons is examined 
at finite chemical potential and at low temperature with Chern-Simons (CS) term which connects the instantons and
the $U(1)$ gauge field \cite{HSSY}. From the model of the nuclear matter given here, we can derive the equation of state (EoS).
Then it can be applied to the compact cold star (like the neutron star)
to estimate the relation of the mass and the radius.


For the deconfined phase (C), up to now, some holographic models have been applied to investigate the neutron star 
with a quark matter core\,\cite{Hoyos,FRE,Fadafan2}. 
On the other hand, the holographic investigation of the matter in the region (B) has not been performed well to study the neutron star, particularly for the mass-radius relation. Of course, several investigations for baryons have been done with the holographic models; see Refs.\,\cite{Nawa:2008uv, Rho:2009ym, Preis:2016fsp, Elliot-Ripley:2016uwb, Ishii:2019gta, Nakas:2020hyo} for examples of previous studies.

In our approach, 
instead of solving the equations of motion,
the flavored gauge fields are replaced by the instanton solution given in
$R^4$ flat space with an instanton of size $\rho$. 
The solution is regarded as a trial function of the solutions of their equations of motion. 
Since, in our case, the instanton is embedded in a deformed 4D space of the curved 6D background space-time, then the size $\rho$ is not arbitrary, and it should be fixed at a suitable value \cite{GKNTT,HSSY}.
This value is determined here by minimizing the free energy of the system at $\rho=\rho_\mathrm{min}$. Then the other physical quantities (the free energy, the chemical potential, and others) are 
obtained by using this $\rho_\mathrm{min}$ and other parameters of the theory.
As a result, the resultant free energy can be expressed as a function of $\mu$ \cite{GKNTT}.
Then through this approach, we can obtain the EoS of the instanton gas;
namely the nuclear matter.

 
Another quantity which characterize the instanton gas is the speed of sound  ($C_s$) of the system.
We find the condition to its value as
$1/2 <C_s^2 <1$ ;
although it preserves the causality since it is smaller than unity, the lower bound is 1/2, 
which is large compared to that of the nuclear matter.


In this article, $SU(2)$ flavored vector fields are introduced. And the instanton-type soliton solution 
of the vector fields is considered as the baryon,
which can be observed in the confined region as shown
in Fig.~\ref{Phase-diagram}.
In the confined phase, the CSC phase is realized for $\mu >4.7$ \cite{GKNTT}.
However, when the back-reaction is considered as in the present case,
this CSC phase region is replaced by the newly appeared RN deconfined phase, which  becomes dominant for $\mu >1.7$ \cite{GKNTT}.
Since we are considering the baryons in the confined phase,
we avoid discussing the EoS in CSC phase. 

\smallskip
In the next section, our bottom-up model is proposed by introducing a dilute gas of instantons.
In Sec.\,\ref{sec:LTconfinement}, we explain how to embed the instantons
in the confined background metric. 
In Sec.\,\ref{sec:action_size}, the free energy of the system is computed according to our method,
and 
EoS for the instanton gas is obtained
at low temperature. In Sec.\,\ref{sec:TOVeq}, using the EoS, the Tolman-Oppenheimer-Volkov (TOV) equations 
for a compact star are solved to find the mass-radius ($M$-$R$) relation. In Sec.\,\ref{sec:comparison}, our results
are compared with the ones
from nuclear physics and astronomical observations.
In the final section, summary and discussions are given.

\section{Bottom-up CSC model with nuclear matter}

We start from the bottom-up 
holographic model which is proposed as YM theory with color superconducting phase. 
It is given by the following 
gravitational theory \cite{GKNTT,Gub,Hart}.
\bea\label{bottom-up}
S &=& \int d^{d+1} \xi \sqrt{-g}\,\mathcal{L}  \,, \label{action-0} \\
 \mathcal{L}&=&\mathcal{L}_\mathrm{Gravity}+\mathcal{L}_\mathrm{CSC} \,,   \label{action-1} \\
     \mathcal{L}_\mathrm{Gravity}&=&  {\cal R} + {d(d-1) \over L^2}\,, \label{bulk-L} \\
     \mathcal{L}_\mathrm{CSC} &=& - {1 \over 4} F^2   - |D_{_\mu} \psi|^2 - m^2 |\psi|^2\, , \label{probe-L} \\
   F_{\mu\nu}&=&\partial_\mu A_\nu-\partial_\nu A_\mu\,,\quad
D_{\mu} \psi = (\partial_{\mu}-iqA_{\mu})\psi\, .
\eea
This describes $(d+1)$-dimensional gravity coupled to a $U(1)$ gauge field, $A_\mu$, 
and a charged scalar field, $\psi$. Here we consider the case of $d=5$.
The charge $q$ denotes the baryon number of the scalar $\psi$, and it is chosen as $q=2/N_c$ to represent the quark-Cooper pair formation.
The mass $m$ is given to reproduce the corresponding conformal dimension of the diquark
operator dual to the scalar field $\psi$ with the charge $q=2/N_c$. 
Here, we put $1/2\kappa_6^2 =1$, and the dimensionful $U(1)$ gauge-coupling
constant is set as unit. So the metric is dimensionless and $A_{\mu}$ has 
dimension one as in the (3+1)-dimensional case.

Previously, the above holographic model has been considered to be dual
to the superconductor of the electric charge \cite{Gub,Hart}
and of the $R$ charge \cite{Nishi}.
And it was recently extended to a theory dual of the color superconductor
in \cite{Basu,GKNTT,Fadafan}.

Here we add the sector of the $SU(N_f)$ gauge fields, which is used to build the nuclear matter through the instanton configuration, which
generates the baryon number in QCD.
Here, the instanton configuration is deformed since it is embedded
in a higher-dimensional curved space.
\bea
   \mathcal{L}&=&\mathcal{L}_\mathrm{Gravity}+\mathcal{L}_\mathrm{CSC} +\mathcal{L}_\mathrm{V} \label{action-V}\,, \\
   \mathcal{L}_\mathrm{V} &=& - {1 \over 4} {\rm tr}F^2_{SU(N_f)} \label{vector-Nf}
\eea
where $F^2_{SU(N_f)}$ denotes the 2-form $SU(N_f)$ gauge field. 
The coupling constant $g_{\rm YM}^2$ is set as one. 
 The mass dimension of the gauge field is taken to be one.
Then, in order to pick up the coupling of
the baryon number current and the instanton configuration, we add the CS term, 
\beq
  S = \int d^{d+1} x \sqrt{-g}\,\mathcal{L} +S_{\rm CS}\,. 
\eeq

We notice that there are two ways to proceed with the analysis
based on this model since both the matter and the gravity fields are living
in the same 6D spacetime.
Therefore, the matter fields are not confined to special $D$-branes:

(i) So, in one way, we can solve the equations of motion of all fields,
gauge fields and the gravity, in the same way. In this case,
the back-reaction of the gauge fields are included in the metric.

(ii) On the other hand, it is possible to restrict our calculation
to the probe approximation where the equation of motion of the matter fields
are solved on a fixed background given by the solution
of the Einstein equations of $\mathcal{L}_{\rm Gravity}$.
This approximation is assured for the large gauge couplings.

Hereafter we proceed with the analysis according
to the probe approximation (ii).
The matter parts are considered as the probe of the system.
So, first we fix the gravitational background
by solving ${\cal L}_{\rm Gravity}$.

\subsection{Instanton in $R^4$}

Before solving the equation of motion of our model, the ansatz for the $U(1)$ gauge field and the instanton configurations for $SU(2)$ gauge fields
are given. 
The coordinates are denoted as
$ (\xi^0,\xi^1,\xi^2,\xi^3,\xi^4,\xi^5)=(x^0,x^1,x^2,x^3,w,z)$.
Then, we set the ansatz for $U(1)$ gauge field as
\beq
   A_b=A_b(z)\delta_b^0\, .
\eeq
The $SU(2)$ vector fields are set as
\beq
   \vec{f}_1= F_{ab}^i\tau_i\, ,
\eeq
and the following ansats are imposed,
\bea
   (\vec{f}_1)_{ij}&=&Q(x^m-a^m,\rho)\epsilon_{ijk}\tau^k\, ,\label{an1} \\
   (\vec{f}_1)_{iz}&=&Q(x^m-a^m,\rho)\tau^i\, ,\label{an2}
\eea
where $\rho$ ($a^m$) denotes the instanton size (position), $\epsilon_{123z}=1$, 
$i,j=1,2,3$ and  $m=1,\dots ,4$, where $x^4=z$.
Then the $SU(N_f)$ vector part (\ref{vector-Nf}) is given for $N_f=2$ as 
\beq
 \mathcal{L}_\mathrm{V} 
                   = -{3\over 2}Q^2\left[\left({g}^{11}\right)^2
        +{g}^{11}{g}^{zz}\right]\ .  
           \label{LV-2}
\eeq

\vspace{.5cm}

In order to see the baryon, $Q$ is given as an instanton solution in flat 4D space $\{x^m\}$
\cite{HSSY}, 
\beq\label{Inst-Sol}
   Q={2\rho^2\over \left((x^m-a^m)^2+\rho^2\right)^2}\, .
\eeq
However, in general, this is not a solution of our bottom-up model since this configuration is embedded in a curved space. 
So, here  we introduce it as a trial function
which is supposed to be a solution of the system under an appropriate condition. The trial function is 
given in our formulation such that
the size
parameter $\rho$ of the instanton configuration is determined to satisfy the variational principle of the
energy density in our bottom-up model. 
We determine it by minimizing the embedded instanton energy
as in Refs.\,\cite{GKTTT,HSSY}. We notice that there is an another method giving $Q$ --- including its $z$ dependence --- by solving the embedding equations
of motion \cite{RSRW}. Here we solve only for the size parameter.

According to the above strategy, we study multi-baryon state by replacing $Q(z)$
by the multi-instanton form with dilute gas approximation. Then we
rewrite $Q$ as
\beq\label{multi-i}
  Q^2=\sum_i^{N_\mathrm{I}}{4\rho^4\over \left((x^m-a^m_i)^2+\rho^2\right)^4}\, ,
\eeq
where the overlap among the instantons are suppressed in obtaining $Q^2$.
This point is checked for our solutions.

In the case of the flat space, in $R^4$,
we find the energy density as a sum of each single instanton,
\bea
  \int d^4\xi^m Q^2&=& 2N_\mathrm{I}\int_0^{\infty} dz \bar{q}(z)^2\, \nonumber \\
  &=&{2\pi^2\over 3}N_\mathrm{I}\, ,
\eea
where
\beq\label{multi-i-2}
  \bar{q}^2={\pi^2\rho^4\over 2(z^2+\rho^2)^{5/2}}\, 
\eeq
for the case of dilute gas, where interactions between instantons are neglected. Then the result
is given by one instanton ``mass'' times its number $N_\mathrm{I}$. 

In the present case, the instantons are embedded in a curved space.
The metrics are functions of the coordinate $z$.
This implies that the state of the baryon is affected dynamically
by the gluons. In other words, the baryon is stable when it is embedded
in the background corresponding to the confined phase.
Furthermore, these baryons would construct a nuclear matter
like the neutron star as we see below.

\vspace{.5cm}
\section{Low temperature confined phase and the fourth coordinate $z$}
\label{sec:LTconfinement}


We suppose the nuclear matter would be the gas of instantons made up
of the $SU(2)$ gauge fields in the confined phase.
The spacetime background solution,
which is dual to the low temperature confined phase,
is given from $\mathcal{L}_\mathrm{Gravity}$ at
low temperature 
as the AdS-soliton solution. 
It is obtained as
\beq\label{Soliton}
   ds^2=r^2(\eta_{\mu\nu}dx^{\mu}dx^{\nu}+f(r)dw^2)+{dr^2\over r^2f(r)}\, ,
\eeq
where 
\beq\label{Soliton-2}
 f(r)=1-\left({r_0\over r}\right)^5\, , \quad r_0={2\over 5R_w}\, ,
\eeq
and $2\pi R_w$ denotes the compactified length of $w$,
which denotes the coordinate of the compact $S^1$.  
We notice that this solution is also a back-reacted one
when the $SU(2)$ gauge fields are set to be zero.
In this case, the phase diagram in the $\mu$-$T$ plane
is given in Ref.\,\cite{GKNTT}.

Here we restrict the region of the parameters, $T$ and $\mu$,
to the confined phase. In fact, under the configuration (\ref{Soliton}),
one finds a linear potential between quark and antiquark
by evaluating the Wilson loop \cite{Witt}. 
Hence the vacuum stays in the confined phase,
and the instanton number is identified with the baryon number.
Then, in this confined phase, we can examine the nuclear matter
through the gas of instantons.

\subsection{Coordinate $z$ and embedding of instanton}

In proceeding the analysis mentioned above, it is necessary to choose the coordinate $z$ carefully. 
For example, we may choose the four coordinates to make the instanton as,
\beq\label{z-r}
 (\xi^0,\xi^1,\xi^2,\xi^3,\xi^4)=(x^0,x^1,x^2,x^3,r)\,, 
\eeq
namely as $z=r$. In this case, using $(\ref{LV-2})$ and $(\ref{Soliton})$,
we have
\bea\label{SU2-gauge}
S_{\rm V} &=& \int d^{6} \xi \sqrt{-g}\,\mathcal{L}_{\rm V}  \,, \label{action-g-2} \nonumber \\
             &=& \int dx^0dx^1dx^2dx^3dw dr~ r^4 \left(-{3\over 2}Q^2\left[{1\over r^4}+f(r) \right] \right)\,  \nonumber \\
            &=& \int dx^0 dwdr~ r^4 \left(-{3\over 2}2N_{\rm I} \bar{q}^2\left[{1\over r^4}+f(r) \right] \right)\,  \nonumber \\
            &=&N_{\rm I} \int dx^0 dwdr~ r^4 \left(-{3\over 2} {\pi^2\rho^4\over (r^2+\rho^2)^{5/2}} \left[{1\over r^4}+f(r) \right] \right)\ .         
\eea
Here we find a logarithmic divergence in the integration over $r$. It is understood as follows:
\bea
      S_{\rm V} & \propto&  \int^{\infty}_{r_0} dr~ r^4 \left(-{3\over 2} {\pi^2\rho^4\over (r^2+\rho^2)^{5/2}} \left[{1\over r^4}+f(r) \right] \right)\,   \nonumber \\
                   & \sim\atop_{ r \to\infty} &  \int^{\infty} dr  \left(-{3\over 2} {\pi^2\rho^4\over r} \right) 
                     \sim  -{3\over 2} {\pi^2\rho^4} \log(\infty) \,.
\eea
This fact implies that we should reset $z$ to other coordinate which leads to a finite energy density of the instanton. 

\vspace{.3cm}
A clue to finding such coordinate is seen in the top-down model
of D4/D8 model \cite{GKNTT,SS,HSSY}.
In the case of the top-down model, D8-flavor brane is embedded
in a set of special coordinates,
which provide a flat space of $r$-$w$ plane near the point $r=r_0$
by avoiding a conical singularity. 
This is performed in the present case by changing the coordinates
from $(r,w)$ to $(z,\theta)$ as,
\bea
   r^5 &=&r_0^5+r_0^3 z^2 \equiv k\,, \\
   \theta &=& {5\over 2} w\,.
\eea
In fact, in this case, the two dimensional part of (\ref{Soliton}) is rewritten as
\bea
   ds^2_{(2)} &=& r^2 f(r)dw^2+{dr^2\over r^2f(r)}\,  \nonumber \\ 
               &=&{4\over 25}\left({r_0\over r}\right)^3\left({dz^2\over r^2}+z^2d\theta^2 \right)\,.
\eea
The second equation shows that the two dimensional metric is the polar coordinate of 2D flat space. 
Hereafter, we take $r_0=1$ for simplicity as in the top-down case.
In this coordinate, the bulk 6D metric
is written as
\beq
   ds^2=k^{2/5} \eta_{\mu\nu}dx^{\mu}dx^{\nu} +{4\over 25} k^{-3/5} (k^{-2/5} dz^2+z^2d\theta^2) \, ,
\eeq 
where $k=1+z^2$.

Here, noticing $\sqrt{-g}=(4/25)z$, the energy of the embedded instantons is given as 
\bea\label{SU2-gauge-2}
S_{\rm V} 
             &=& \int dx^0dx^1dx^2dx^3 dz~ {4\over 25} z \left(-{3\over 2}Q^2\left[k^{-4/5}+{25\over 4}k^{3/5} \right] \right)\, 
             \nonumber \\
            &=& N_{\rm I} \int dx^0 dz~ {4\over 25} z \left(-{3\over 2} {\pi^2\rho^4\over (z^2+\rho^2)^{5/2}}  \left[k^{-4/5}+{25\over 4}k^{3/5} \right] \right)\, ,  
\eea
where the factor $\int\!d\theta =2\pi$ is absorbed to the gauge coupling constant. 
We can assure the finiteness of the $z$ integration of the above action.

\subsection{CS term} 
The baryon number is given by 
\beq
 N_\mathrm{B}={1\over 32\pi^2}\int d^3x dz \epsilon^{m_1\cdots m_4}{\rm tr}(F_{m_1m_2}F_{m_3m_4})\, .
\eeq
Then the coupling of $A_0$ to the baryon number is given by the following CS term,
\bea
      S_{\rm CS} &=&  
     \kappa_{\rm CS}
\epsilon^{m_1\cdots m_4} \int
  d^4x~dz~ A_0{\rm tr}(F_{m_1m_2}F_{m_3m_4})\,  \nonumber \\
      &=& 24\kappa_{\rm CS}  \int d^4xdz~ A_0 Q^2\, \nonumber  \\
      &=& 48 n \kappa_{\rm CS}  \int d^4xdz ~A_0\bar{q}^2\, ,
\eea
where 
\beq
n={N_{\rm I}\over V_3}\,, \quad V_3=\int d^3x=\int dx^1dx^2dx^3.
\eeq

\section{Effective action and the size of the instanton}
\label{sec:action_size}

The matter action with embedded instantons is obtained here as
\bea\label{bottom-up-3}
 S_{\rm matter}&=& \int d^{6} \xi \sqrt{-g}\, \left( - {1 \over 4} F^2 - {1 \over 4} {\rm tr}F^2_{SU(2)}\,\right)+S_{\rm CS}\, 
 \nonumber \\
                   &=&  
      \int d^4x dz  \left( {1\over 2} z k^{3/5} {A_0'}~^2 - nz{12\over 25}\bar{q}^2\left[k^{-4/5}+{25\over 4}k^{3/5} \right]+n n_0A_0 \bar{q}^2  \right)\,,
  \label{a-m}
\eea
where $A_0'=\partial_z A_0(z)$, 
and $n_0=48 \kappa_{\rm CS}$. 
In order to estimate this action, we solve the equation of motion of $A_0(z)$. It is obtained as
\beq\label{eq-A0}
  -\partial_z\left( zk^{3/5}A_0'\right) +nn_0\bar{q}^2=0 \,.
\eeq
Then we have
\beq\label{eq-A1}
       zk^{3/5}A_0' = \bar{d}={\pi^2\over 6}nn_0{2z^3+3z\rho^2 \over (z^2+\rho^2)^{3/2}} +c \, .
\eeq
Here, $c$ denotes an integration constant of Eq.\,(\ref{eq-A0}) over $z$. We take $c=0$ to set as
$\bar{d}(z=0)=0$. Due to this condition, 
the matter action, $S_{\rm matter}$ of (\ref{a-m}), is written as
\beq
 S_{\rm matter} =  \int d^{4} x \left\{ \mu \bar{Q}-\int dz \left(
{\bar{d}~^2\over 2 z k^{3/5} } + nz{12\over 25}\bar{q}^2\left[k^{-4/5}+{25\over 4}k^{3/5} \right]\right) \right\}\,, 
\eeq
where the chemical potential $\mu$ and the charge density $\bar{Q}\equiv {\bar{d}(\infty)}$ are given as
\bea                     
       \mu &=& A_0(\infty) =  \int_{z_1}^{\infty} dz {\bar{d}\over zk^{3/5}} +A_0(z_1)\, \nonumber  \\
             &=& \int_{z_1}^{\infty} dz {\bar{d}\over zk^{3/5}}    \, ,  \label{chemi} \\
      \bar{Q} &=& \bar{d} (\infty) = {\pi^2\over 3} nn_0 \,,
\eea
where $z_1$ is an arbitrary positive value and $A_0(z_1)$ is set as 
\beq
  A_0(z_1)=0\,,
\eeq
which is the boundary condition in solving the differential equation of $A_0(z)$, Eq.\,(\ref{eq-A1}). \\
Then the free energy density $ {\cal E}$ of the instanton system is given as 
\bea
   S_{\rm matter}   &=& -\int d^5\xi ~~ {\cal E}(\rho,\mu)\,   \\
         {\cal E}(\rho,\mu) &=& \int dz \left(
{\bar{d}~^2\over 2 z k^{3/5} } + nz{12\over 25}\bar{q}^2\left[k^{-4/5}+{25\over 4}k^{3/5} \right]\right) - \mu \bar{Q} \, \label{GFE}  
\eea

\subsection{Size of instanton}

The free energy density $ {\cal E}(\rho,\mu)$ in (\ref{GFE}) is given as a function of $\rho$ for fixed $n, n_0, z_1$. Then we find the value of $\rho=\rho_{\rm min}$ where 
 $ {\cal E}(\rho,\mu(\rho))$ takes its minimum for a set of $(n, n_0, z_1)$;
for this $\rho_{\rm min}$, $\mu$ and $ {\cal E}(\rho,\mu)$ are determined.
Repeating this procedure by changing only the value $n$ of the set $(n, n_0, z_1)$, we find different $\rho_{\rm min}$.
As a result, we obtain the relation between $\mu$ and $ {\cal E}(\rho,\mu)$.  

We notice that the results obtained in this way depend on the other two parameters $(n_0, z_1)$. 
We give a comment related to the parameter $z_1$.
For $z_1=0$, we have a problem that
the values of $\mu$ and  ${\cal E}$ at $\rho=0$ are divergent,
\bea
  \mu|_{\rho\to 0}&=& \int_0^{\infty} dz {\bar{d}(\infty)\over zk^{3/5}} \,  \nonumber \\
              &\sim\atop_{ z \to 0}& \bar{d}(\infty) \int_0{dz\over z} \to +\infty\,,  \\
             {\cal E}(\rho=0, \mu)| &\sim \atop_{ z \to 0} & \int_0 dz \left(
                       -{ \bar{d}(\infty)^2 \over 2 z k^{3/5} } \right) \to -\infty .
 \eea

In the present model these divergences are evaded by setting $z_1$, the lower bound of $z$, at a finite value.
On the other hand, in the case of the top-down model \cite{GKTTT}, there are no such divergences.
In fact, we can see that $\mu$ is finite even if $\rho$ is zero for $z_1=0$.
Therefore, we expect the existence of 
some improvement
of the present bottom-up model to resolve the above divergences. 
Here, this point remains as an open problem,
 and $z_1$ is introduced as a simple cutoff parameter for the above undesirable infrared divergences.


\subsection{ EoS for nuclear matter}

By giving $n, n_0, z_1$,
(which are the parameters of the present bottom-up model),
$\mu(\rho)$ and ${\cal E}(\rho)$ are calculated according to the Eqs.\,(\ref{chemi}) and (\ref{GFE}) as the functions of $\rho$. Then 
a minimum point of ${\cal E}(\rho)$ is found at $\rho=\rho_{\rm min}$. Thus we can obtain ${\cal E}(\rho_{\rm min})$ and $\mu(\rho_{\rm min})$
for the given parameters $n, n_0, z_1$. 

Repeating this procedure by changing only $n$ with $(n_0, z_1)$ fixed,
then we get the relationship of $\mu$ and ${\cal E}$.
From this relationship we obtain the EoS of the instanton gas
as mentioned above. 

\begin{table}[htbp]
\caption{EoS of nuclear system at low temperature for $n_0=1.5, z_1=0.1$,
 and} $v_{\rm min}={4\over 3}\pi\rho_{\rm min}^3 n$.
\begin{center}
\begin{tabular}{ccccc}
$n$ & $\mu(\rho_{\rm min})$ &  ${\cal E}(\rho_{\rm min})$ & $\rho_{\rm min}$ & $v_{\rm min}$ \\
\hline 
0.005 & 0.0688 &0.00031 & 0.01 & 2.09 $\times 10^{-8}$ \\
0.01 & 0.138 & 0.0000393 & 0.02 & 3.35$\times 10^{-7}$ \\
0.015 & 0.209 &-0.00022 & 0.03 & 1.70$\times 10^{-6}$ \\
0.02 & 0.275 & -0.00178 & 0.04 & 5.36 $\times 10^{-6}$ \\
0.03 & 0.411 & -0.00829 & 0.06 & 2.75 $\times 10^{-5}$ \\
0.05 & 0.68 & -0.0368 & 0.09 & 1.52$\times 10^{-4}$ \\
0.1 & 1.335 & -0.2075 & 0.13 & 9.2 $\times 10^{-4}$ \\
\hline
\end{tabular}\label{table-1}
\end{center}
\end{table}



\vspace{.3cm}
In Table \ref{table-1}, we show a resultant example of such calculations for $n_0=1.5, z_1=0.1$.
The regions of the calculations are restricted to the region for $\mu <1.7$ \cite{GKNTT}.
On the other hand, the pressure $p$, which is given by
$p=-{\cal E}$, of the instanton gas is negative for $\mu<\mu_c\sim 0.17$. 
So in this region, the gas is in an undesirable phase as a nuclear matter considered here. 
\footnote{The negative pressure state might be constructed under a delicate balance of two kinds
nuclear forces \cite{Ber, CLY}. Although this point is very interesting, we will discuss it in future work.}

Thus we find that the stable nuclear matter exists in the region, $1.7> \mu>\mu_c\sim 0.17$. 
In this region of the nuclear matter, 
its dilute gas
picture is reasonable since $v_{\rm min}\sim O(10^{-4}) \ll 1$, where
$v_{\rm min}={4\over 3}\pi\rho_{\rm min}^3 n$ indicates the volume which is occupied by instantons in a unit 3D volume.

\begin{figure}[htbp]
\begin{center}
\includegraphics[width=14.0cm,height=6cm]{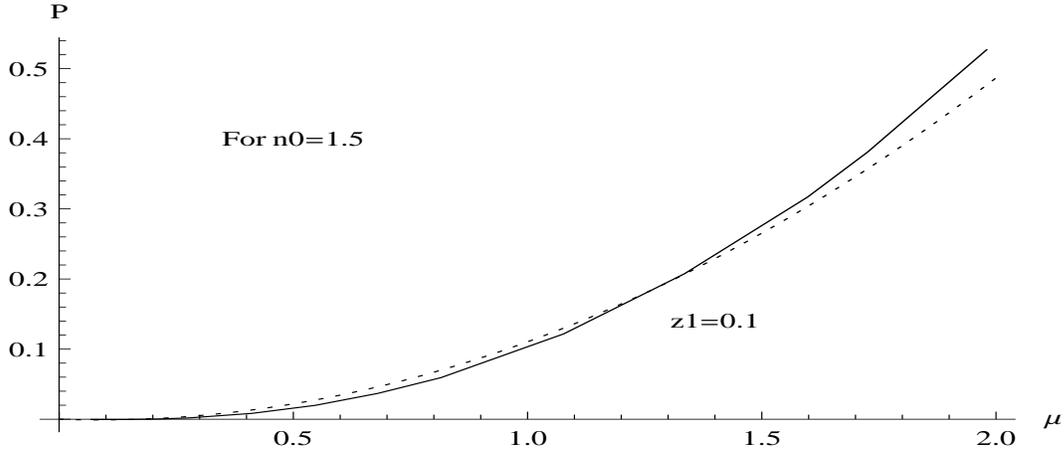}
\caption{{\small  $ p=-{\cal E}(\mu)$ versus $\mu$ at stable instanton size
 $\rho_{\rm min}$ near phase transition point $\mu_c = 0.17$. The solid curve denotes the
numerical calculations, and the dotted curve represents $p=0.13 \mu(\mu-\mu_c)$.
}}\label{delta-E2}
\end{center}
\end{figure}

\vspace{.3cm}
Here we proceed the analysis, 
and we can arrive at the following approximate formula 
\beq
  p=a\mu(\mu-\mu_c)\,, \label{pressure}
\eeq
where $a=0.13$ and $\mu_c=0.17$ (See Fig.~\ref{delta-E2}). 
Then using (\ref{pressure}).
the energy density is given at $T=0$ as \cite{FRE}
\beq\label{energy}
   {\epsilon} = \mu q-p=\mu{\partial p\over \partial\mu} -p =a\mu^2\,,
\eeq
and the speed of sound is obtained as
\beq
    C_s^2 = {\partial p\over \partial \epsilon}={\partial p/\partial \mu \over \partial \epsilon/\partial \mu}  
           = 1-{\mu_c\over 2\mu}\, .
\eeq

This leads to the following bound of $C_s^2$, 
\beq\label{cs}
     {1\over 2} <C_s^2<1\, .
\eeq
We notice that the lower bound $1/2$ is fairly large comparing to that of the ordinary nuclear matter. 
We should however notice that the constraint (\ref{cs}) for $C_s^2$ comes from the formula (\ref{pressure}), which is an approximate formula available at small $\mu$.
It is possible to improve (\ref{pressure}) by adding correction terms of higher powers of $\mu$. For example, by adding a term like $\mu^4$, we can find a new formula,
$p=a'\mu(\mu-\mu_c')+b'\mu^4$ with appropriate values for the parameters, $a', \mu_c', b'$. In this case, we can see that the maximum value of $C_s$ is realized at $\mu$
smaller than 1 and its maximum value is
suppressed from 1, then it decreases to $C_s=1/\sqrt{3}$ in the large $\mu$ region. 
However, as we can see, the correction term of $\mu^4$ becomes small 
in the small $\mu$ region. Then it is difficult to suppress largely the lower bound 1/2. We find a small value of sound speed in region of $0<C_s^2<1/2$ for
$\mu<\mu_c$, where $\mu_c(\neq \mu_c')$ denotes the critical point and it satisfies $p(\mu_c)=0$. 
In this region of $\mu$, however, the pressure $p$ is negative, then the instanton gas cannot make a nucleon matter like a star.
This implies that our model cannot cover the low-density part of the normal nuclear matter, in which the sound speed decreases to zero from
its upper bound $(1/\sqrt{3})$. How to overcome this point remains
as a future problem.

In this sense, the nuclear matter
given here might be a special one.
Our model is considered in a restricted region of density or $\mu$, $\mu_c<\mu<1.7$, where 1.7 denotes the transition point to the deconfined
RN phase. Therefore, we continue our analysis by using a simple model with
(\ref{pressure}) and (\ref{energy}), then we arrive at the EoS of the nuclear matter given as the instanton gas. It is
written as,
\beq\label{eos}
  p=\epsilon-\sqrt{a\epsilon}\,\mu_c\, .
\eeq

\newcommand{\pd}[3]{\left(\frac{\partial#1}{\partial#2}\right)_{\!#3}}

\subsection{Alternative approach to find EoS}


In the present scheme, the free energy density (\ref{GFE}) takes $n$,
$n_0$, $\rho$, and $z_1$ as independent variables.
Among them, $n_0$ is an intrinsic parameter of the theory and $\rho$ is
determined dynamically such that $\mathcal{E}$ takes its minimum value,
while $z_1$ is set to be a plausible value for the moment.
The free energy density is expanded in a power series of $n$,
\begin{equation}
\mathcal{E}(n,\rho,n_0,z_1)=n\,c_1(\rho)-n^2n_0^2\,c_2(\rho,z_1)\ ,
\end{equation}
in which
\begin{eqnarray}
c_1(\rho)
&=&\frac{12}{25}\frac{\pi^2\rho^4}{2}\int_0^\infty\!\!dz\,
\frac{z}{(\rho^2+z^2)^{5/2}}
\left[(1+z^2)^{-4/5}+\frac{25}{4}(1+z^2)^{3/5}\right]\ ,\\
c_2(\rho,z_1)
&=&\left(\frac{\pi^2}{3}\right)^2\int_0^\infty\!\!dz\,
\frac{1}{(1+z^2)^{3/5}}\left[\frac{1}{2}\frac{2z^2+3\rho^2}{(\rho^2+z^2)^{3/2}}
-\frac{1}{8}\frac{z(2z^2+3\rho^2)^2}{(\rho^2+z^2)^3}\right]\nonumber\\
&&\hspace{0.5in}
-\left(\frac{\pi^2}{3}\right)^2\!\int_0^{z_1}\!dz\,
\frac{1}{(1+z^2)^{3/5}}\,\frac{1}{2}\,\frac{2z^2+3\rho^2}{(\rho^2+z^2)^{3/2}}
\ .
\end{eqnarray}

In order to change the independent variable from $n$ to $\mu$,
we notice that $\mu$ can be written as
\begin{equation}
\mu(n,\rho,n_0,z_1)=nn_0\,h(\rho,z_1)\label{eq:mubyh}
\end{equation}
with
\begin{equation}
h(\rho,z_1)=\frac{\pi^2}{6}\,\int_{z_1}^\infty\!dz\,
\frac{1}{(1+z^2)^{3/5}}\,\frac{2z^2+3\rho^2}{(\rho^2+z^2)^{3/2}}\ .
\end{equation}

Now, eliminating $n$ by (\ref{eq:mubyh}),
the free energy density is rewritten as
\begin{equation}
\mathcal{E}=
-\frac{c_2(\rho,z_1)}{(h(\rho,z_1))^2}\,\mu^2
+\frac{c_1(\rho)}{n_0h(\rho,z_1)}\,\mu\ .
\label{eq:ab_coeff}
\end{equation}
From Eq.~(\ref{eq:ab_coeff}) 
one finds that the coefficients ($a$ and $b$) and
the critical chemical potential ($\mu_c$) are given by
\begin{equation}
a=\frac{c_2(\rho,z_1)}{h(\rho,z_1)^2}\ ,\qquad
b=\frac{c_1(\rho)}{n_0\,h(\rho,z_1)}\ ,\qquad
\mu_c=\frac{c_1(\rho)\,h(\rho,z_1)}{n_0\,c_2(\rho,z_1)}\ .
\end{equation}

Given $\mu$, $z_1$, and $n_0$, the minimum point of $\mathcal{E}$
({\textit i.e.\/}, the maximum point of $p$) in the $\rho$ space
can be sought numerically.



The minimum energy point depends on the variable which is taken fixed.
Actually, two types of differential coefficients have different values as
\begin{equation}
\pd{\mathcal{E}}{\rho}{n}-\pd{\mathcal{E}}{\rho}{\mu}
=\pd{\mathcal{E}}{\mu}{\rho}\pd{\mu}{\rho}{n}
=-\pd{\mathcal{E}}{n}{\rho}\pd{n}{\rho}{\mu}\ .
\end{equation}
At the point $\rho=\rho_\mathrm{min}$
such that $(\partial\mathcal{E}/\partial\rho)_n=0$,
it holds that
\begin{equation}
\pd{\mathcal{E}}{\rho}{\mu}
=-\pd{\mathcal{E}}{\mu}{\rho}\pd{\mu}{\rho}{n}
=(-b+2a\mu)nn_0\,h_\rho(\rho,z_1)\ .\label{eq:sign-eval}
\end{equation}
In (\ref{eq:sign-eval}), one finds that
\begin{eqnarray}
h_\rho(\rho,z_1)
~\equiv~\frac{\partial h(\rho,z_1)}{\partial\rho}
&=&-\frac{\pi^2}{2(2a_0)}\int_{z_1}^\infty\!dz\,
\frac{1}{(1+z^2)^{3/5}}\,\frac{\rho^3}{(\rho^2+z^2)^{5/2}}<0\ ,\\
-b+2a\mu
&=&a(\mu-\mu_c)+a\mu>0\ ,
\end{eqnarray}
in the present situation, \textit{i.e.\/}, $a>0$ and $\mu>\mu_c$.
Therefore, we conclude that
\begin{equation}
\pd{\mathcal{E}}{\rho}{\mu}<0\quad\hbox{(at $\rho=\rho_\mathrm{min}$)}\ .
\end{equation}
This means that the minimum energy point along an $n$-fixed curve
has much lower energy points towards the $\rho$-increasing direction
with $\mu$ being kept constant.
%
However, the problem is whether the minimum energy significantly depends
on the fixed variable or not.

To estimate the difference between the two types of minimum energies, one should continue numerical calculations on $\mathcal{E}$ from $\rho_\mathrm{min}$ by fixing $\mu$ and by increasing $\rho$ until one finds another minimum energy density $\mathcal{E}'_\mathrm{min}$.
In the range of $0.015\leq\rho\leq 0.1$ according to Table~\ref{table-1},
the calculations make the difference of the minimum energies explicit
quantitatively such as
$(\mathcal{E}'_\mathrm{min}-\mathcal{E}_\mathrm{min})/\mathcal{E}_\mathrm{min}
\lesssim 0.028$.

We may consider that such differences do not affect characteristic features of
EoS, and we again employ $n$ as the independent variable of $\mathcal{E}$
in the succeeding sections.


\section{TOV equation and $M$-$R$ relation of neutron star}
\label{sec:TOVeq}
 
The TOV equations for a star with the mass $m$, and $p$ at the radius $r$
 in the star are given as,
\bea
   {dp\over dr} &=& -G(\epsilon+p){m+4\pi r^3p \over r(r-2Gm)}\,,  \label{TOV-1}\\
   {dm \over dr} &=& 4\pi r^2\epsilon\,.   \label{TOV-2}
\eea
Here, $G$ denotes the gravitational constant. They can be solved by using the EoS given in (\ref{eos})
with the boundary condition,
$p(r=0)=p_c$ and $m(r=0)=0$. 
Solving these equations, we obtain the mass of the star, $M=m(R)$, and its radius $R$, where 
$R$ is defined by $p(R)=0$.


\begin{figure}[htbp]
\begin{center}
\includegraphics[width=14.0cm,height=6cm]{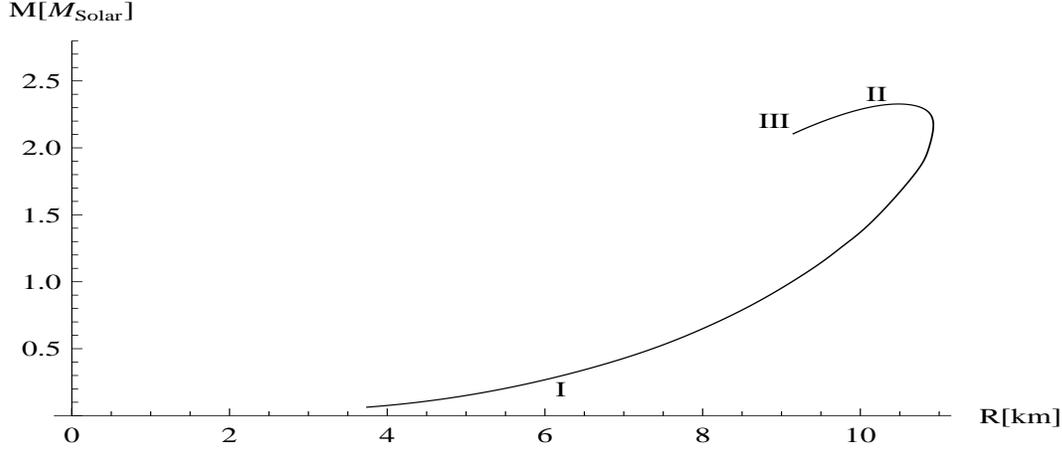}
\caption{{\small  Plot of the neutron mass $ M(R)$ (in unit of solar mass $M_{\odot}$) versus the radius $R$ 
for 
$p=\epsilon-\sqrt{a\epsilon}\mu_c$ (or
$p=a \mu(\mu-\mu_c)$), $\mu_c=0.17$ and $a=0.13$.
}}\label{MR-1}
\end{center}
\end{figure}

In order to obtain the numerical results with dimensionful quantities, we must adjust the scale parameters for
each dimensionful quantities ($\tilde x$).
To do so, the above TOV equations are rewritten 
by using the dimensionless quantities as \cite{Fadafan2}
\bea
   {d{\tilde p}\over d{\tilde r}} &=& -B({\tilde \epsilon}+{\tilde p}){{\tilde m}+4\pi A{\tilde r}^3{\tilde p} \over {\tilde r}({\tilde r}-2B {\tilde m})}\,,  \label{TOV-3}\\
   {d{\tilde m} \over d{\tilde r}} &=& 4\pi A {\tilde r}^2{\tilde \epsilon}\,,  \label{TOV-4} 
\eea
where 
\beq
  A={r_0^3\epsilon_0 \over m_0}\,, \quad B={G m_0 \epsilon_0 \over p_0 r_0},
\eeq
and the various variables ($x$) are replaced by the dimensionless quantities ($\tilde x$) by using its typical dimensionful value
$x_0$ as $x=x_0 {\tilde x}$. For example,
we rewrite as $p=p_0 {\tilde p}$. 


\vspace{.3cm}
Here we give a comment on the replacement, $\epsilon=\epsilon_0 \tilde{\epsilon}$ and how $\epsilon_0$ is determined. 
In order to solve the TOV equation, $\epsilon(r)$ in the equation
is replaced using the Eq.\,(\ref{eos}) as a function of $p(r)$,
\beq
  \epsilon={a\mu_c^2\over 4}\left(1+\sqrt{1+{4p\over a\mu_c^2}}\right)^2\, .
\eeq
Similar to the above, we introduce the dimensionless quantities as
$a=a_0\tilde{a}$, $\mu_c=\mu_0\tilde{\mu}_c$, and
\beq
   \epsilon=\epsilon_0\tilde{\epsilon}\,, 
\eeq
by imposing reasonable conditions
\beq
    \quad \epsilon_0=p_0=a_0\mu_0^2\,.
\eeq
Then in Eqs.\,(\ref{TOV-3}) and (\ref{TOV-4}), we find
\beq
  \tilde\epsilon={\tilde{a}\tilde{\mu}_c^2\over 4}\left(1+\sqrt{1+{4\tilde{p}\over \tilde{a}\tilde{\mu}_c^2}}\right)^2\, ,
\eeq
where
we put the dimensionless number as $\tilde{a}\tilde{\mu}_c^2=0.13\times 0.17^2=3.76\times 10^{-3}$.

Then the solutions of (\ref{TOV-3}) and  (\ref{TOV-4}) with $A=B=1$ are equal to the one of  (\ref{TOV-1}) and  (\ref{TOV-2}) with $G=1$.
Then the solution of the latter equations are translated to the one of the former ones as shown in Fig.\,\ref{MR-1};
it is given for $r_0=3.0$ km 
in natural units. In this case, we find $m_0=2.03 M_\odot $
\footnote{The symbol $M_\odot$ denotes the solar mass. 
}
and $\epsilon_0^{1/4}=0.896$ GeV, 
which provides physical units to the quantities given in the Table 1. For example, the energy density of the center of the star, whose mass is about two solar mass, is given by 0.00242 GeV$^4$. This is about twice as large as the one of the normal nuclear matter.

We can see that the radius and the mass increase with increasing $p_c$, the central value of the pressure. 
The resultant curve rotates anticlockwise toward the smaller radius in the large mass region. 
However from the point II to the point III in Fig.\,\ref{MR-1}, the state would be unstable since $\partial M(\epsilon_c)/\partial \epsilon_c <0$
\cite{Fadafan2} .
On the other hand, in the region from I to II, we can see the behavior $\partial M(\epsilon_c)/\partial \epsilon_c >0$.
Thus, the 
results given here indicate the possibility of the existence of a star with $(M,R)\sim(2M_{\odot}, 10\ \mathrm{km})$. 


\section{Comparison with the observational data}
\label{sec:comparison}

There are some important observational data which can constrain the theoretical EoS.
We now have the following at least three important constraints for the $M$-$R$ relation;
\begin{enumerate}
    \item Two solar mass ($2 M_\odot$) neutron star observation from the Shapiro delay measurement~\cite{Demorest:2010bx,Antoniadis:2013pzd}.
    \item Radius constraint from GW170817 via the gravitational wave observation.
    The actual constraint is $9.0~\mathrm{km} < R < 13.6~\mathrm{km}$ 
for $M=1.4 M_\odot$~\cite{Annala:2017llu,De:2018uhw,Tews:2018iwm}.
    \item The upper bound of $M$ for cold spherical neutron stars which is estimated from no detection of relativistic optical counterpart in the analysis of GW170817.
    The actual limit is estimated as the range $2.15M_\odot-2.26M_\odot$~\cite{Shibata:2017xdx}.
\end{enumerate}
To support $2M_\odot$ neutron stars in the theoretical calculation, we need a sufficiently stiff EoS and it is closely related to the large value of the speed of sound.
Then, the pressure of matter in the inner neutron star core should be large.
From the second constraint -- the radius is relatively small in the moderate $M$ region -- it requires that the pressure of the matter in the outer neutron core should be relatively small.
It is interesting that our bottom-up model is quite simple, but our EoS can manifest the $2M_\odot$ constraint and the restriction that the $M$-$R$ curve must go through $9.0 < R < 13.6$ around $M\sim 1.4 M_\odot$.
However, too stiff EoS is not preferable because of the third constraint.
Unfortunately, the maximum mass $M_\mathrm{max}$ overshoots
 the $2.15M_\odot-2.26M_\odot$ constraint, but we can expect that there are some phase transitions at high density such as the chiral, deconfined, and color superconducting phase transitions which we do not consider in this study.
Then, EoS can be softer than the present one at high density and thus $M_\mathrm{max}$ should be smaller than the present value.
Particularly, EoS becomes soft if there is a first-order phase transition; for example, see Ref.\,\cite{Kojo:2020krb}.

It should be noted that the above constrains can restrict the large and moderate $M$ regions, but the lower $M$ region is not restricted so much.
Of course, there are several nuclear theoretical EoS and thus we can qualitatively discuss the lower $M$ region where the low density part of EoS dominates the $M$-$R$ curve.
With our EoS, the $M$-$R$ curve flows down to the bottom left, but several $M$-$R$ curves with different EoS flow down to the bottom right.
This result may be induced from the fact that our $C_s^2$ cannot be less than $1/2$ even in the sufficiently low-density region.
Such behavior is similar to the $M$-$R$ curves obtained with EoS proposed in Ref.\,\cite{Prakash:1995uw}; the actual behavior of the curves can be seen in Refs.\,\cite{Lattimer:2000nx,Ozel:2015fia} denoted as
``SQM1-3" and 
these EoS which contain the quark matter make self-bounded neutron stars which do not have minimum masses.
The tendency of the $M$-$R$ curves in the small $M$ region in our model may be modified when we suitably introduce ``interactions" to the present model because we employ the dilute instanton gas approximation; the baryonic contribution can be expected to have relatively strong effects on the EoS at low density.
In addition, we here use the dilute gas of the instanton of the $SU(2)$ gauge field and thus the number of the flavor is considered as two; the system contains the up and down quarks and thus there is no strange quark.
This may introduce the unclearness to the present $M$-$R$ curves because the EoS will be softer than the present one if hyperons appear in the system;
since the core of the neutron star is very dense, the hyperon degree of freedom can join the game as the baryonic mode in addition to nucleons.
This softening problem induced by the hyperons is the so-called hyperon puzzle which is a long standing problem in the study of the neuron stars with EoS; see Ref.\,\cite{Weise:2019mou} as an example.
To discuss this problem in the present bottom-up holographic model, we must consider differences between the light quarks and the strange quark, interactions and bound states more deeply.
In the present study, we consider the simple symmetric nuclear matter because the difference between the flavors and effects of the electrons are difficult to include in the model at present.
Therefore, the inclusion of the effects such as the beta equilibrium and charge neutrality are left for our future work.


\section{Summary and discussions}

 In this paper, we have studied cold nuclear matter and its equation of state (EoS) based on the six-dimensional holographic
model, which was investigated  previously to realize the color superconducting phase of QCD. 
The nuclear matter was introduced as a dilute gas of deformed instantons in the AdS soliton background. 
The instantons are electric-charge neutral and they are made up of the $SU(2)$ gauge fields; the action includes the Chern-Simons term
as well as the kinetic term. Owing to the energy balance between these two terms in the action, the size of the instanton
can be determined for fixed parameters of the model. As the result, the EoS for nuclear matter,
\textit{i.e.\/}, pressure as a function of energy density, is obtained. Through the evaluation of the instanton size, it turns out that the
number of instantons in a unit volume is very small so that our dilute gas picture is consistent. Furthermore, there exists
a critical value for the chemical potential below which the pressure becomes negative. 
 The EoS obtained here is very stiff because the speed of sound
$C_s^2>\frac{1}{2}$. As discussed in the Sec. 4.2, this constraint is reduced to our simplified form of the pressure, $p(\mu)$. The lower bound 1/2 can be suppressed
when the functional form of $p(\mu)$ is modified. 

Here, we however consider an EoS which is obtained from the simplified $p(\mu)$ to see the characteristic properties of our model.
Then, we applied the 
simplified
EoS to solve
the Tolman-Oppenheimer-Volkov equations for a compact star numerically and found the mass-radius ($M$-$R$)
relationship. The curve is somewhat similar to the one for strange quark matter in Refs.\,\cite{Prakash:1995uw, Ozel:2015fia}
and we provided a certain interpretation for that.
 Our result is compared with the observational data and it is seen that
the maximum mass overshoots the current constraint since our EoS for the nuclear matter is very stiff.
This might suggest that we have more phases such as quark matter, which softens the current EoS,
 than that of the nuclear matter.
As we have shown in the Sec. 4, our dilute gas picture is appropriate.
There is a reason why we could achieve two solar mass neutron star in spite of the diluteness of the instanton gas.
In our approach, it might be possible that the mass of an instanton is very heavy compared with an ordinary nucleon, which is a specific property in holographic QCD since the nucleon mass is proportional to the number of colors $N_c$. So far we have no definite answer and this is a future issue.

In order to improve our current study, several ingredients are taken into account. One is to go beyond 
the dilute gas approximation of instantons. By doing this, our EoS is modified and eventually the $M$-$R$ curve might be
changed so as to be more consistent with the observational data.
 The other is to study whether the system enjoys
the phase transition from nuclear matter to (perhaps color superconducting) quark matter at higher baryon density.
If such a phase transition is present, the EoS gets soften and the resultant $M$-$R$ curve could be modified.
Furthermore, it will be interesting to extend our current study into the case with hyperon degrees of freedom. To this end,
the holographic treatment of heavy-light meson system (for instance, see \cite{GEK:2007}) must be instructive.
These issues will be considered in the future.
 



\vspace{.3cm}
\section*{Acknowledgments}
We are grateful D. Blaschke and A. Schmitt for their useful comments during the online conference "A Virtual Tribute to Quark Confinement and the Hadron Spectrum" (vConf2021). It is also a pleasure to thank Masayuki Matsuzaki for useful discussions on the neutron stars and TOV equations.


\newpage

\newpage


\begin{thebibliography}{99}

\bibitem{Basu}
 Pallab Basu, Fernando Nogueira, Moshe Rozali, Jared B. Stang, Mark Van Raamsdonk, 
``Towards A Holographic Model of Color Superconductivity'', 
New J.Phys.13:055001,2011, [arXiv:1101.4042 [hep-th]].

\bibitem{GKNTT} 
Kazuo Ghoroku, Kouji Kashiwa, Yoshimasa Nakano, Motoi Tachibana, Fumihiko Toyoda
``Color Superconductivity in Holographic SYM Theory'', 	Phys. Rev. D 99, 106011 (2019), arXiv:1902.01093[hep-th]:

\bibitem{Gub} 
Steven S. Gubser, ``Breaking an Abelian gauge symmetry near a black hole horizon'',  
 Phys.Rev.D78:065034,2008, arXiv:0801.2977 [hep-th].

\bibitem{Hart}
 Sean A. Hartnoll, Christopher P. Herzog, Gary T. Horowitz,
``Building an AdS/CFT superconductor'', 
 Phys. Rev. Lett. 101.031601, [Xiv:0803.3295 [hep-th]];

\bibitem{Hart2}
  S.~A.~Hartnoll, C.~P.~Herzog and G.~T.~Horowitz,
  ``Holographic Superconductors,''
  JHEP {\bf 0812}, 015 (2008)
  [arXiv:0810.1563 [hep-th]].

\bibitem{Nishi}
Tatsuma Nishioka, Shinsei Ryu, Tadashi Takayanagi,
``Holographic Superconductor/Insulator Transition at Zero Temperature'',  
JHEP 1003:131,2010, JHEP03(2010)131, [arXiv:0911.0962 [hep-th]].


\bibitem{Iqbal:2010eh}
  N.~Iqbal, H.~Liu, M.~Mezei {\it et al.},
  ``Quantum phase transitions in holographic models of magnetism and superconductors,''
  Phys.\ Rev.\  {\bf D82}, 045002 (2010).
  [arXiv:1003.0010 [hep-th]].

\bibitem{Cham} 
 Andrew Chamblin, Roberto Emparan, Clifford V. Johnson, Robert C. Myers,
``Charged AdS Black Holes and Catastrophic Holography'',
10.1103/PhysRevD.60.064018, hep-th/9902170:

\bibitem{Fadafan}
Kazem Bitaghsir Fadafan, Jesus Cruz Rojas, Nick Evans, 
``A Holographic Description of Colour Superconductivity'',  
Phys. Rev. D 98, 066010 (2018), [arXiv:1803.03107 [hep-ph]].
 
\bibitem{Nam}
Cao H. Nam, ``A more realistic holographic model of color superconductivity
with higher derivative corrections'',
 arXiv:2101.00882[hep-th]


\bibitem{Bergman} 
Oren Bergman, Gilad Lifschytz, Matthew Lippert, ``Holographic Nuclear Physics'',
JHEP0711:056,2007, arrXiv:0708.0326 [hep-th] : 

\bibitem{RSRW} 
  M.~Rozali, H.~-H.~Shieh, M.~Van Raamsdonk and J.~Wu,
  ``Cold Nuclear Matter In Holographic QCD,''
  JHEP {\bf 0801}, 053 (2008)
  [arXiv:0708.1322 [hep-th]].

\bibitem{GKTTT}
Kazuo Ghoroku, Kouki Kubo, Motoi Tachibana, Tomoki Taminato, Fumihiko Toyoda,
``Holographic cold nuclear matter as dilute instanton gas'', PhysRevD.87.066006,
arXiv:1211.2499 [hep-th].

 \bibitem{SS}
   T.~Sakai and S.~Sugimoto,
   ``Low energy hadron physics in holographic QCD,''
   Prog.\ Theor.\ Phys.\  {\bf 113}, 843 (2005)
   [hep-th/0412141].   

 \bibitem{HSSY}
   H.~Hata, T.~Sakai, S.~Sugimoto and S.~Yamato,
   ``Baryons from instantons in holographic QCD,''
   Prog.\ Theor.\ Phys.\  {\bf 117}, 1157 (2007)   [hep-th/0701280 [HEP-TH]].

\bibitem{Schmitt}
Florian Preis and Andreas Schmitt, ``Layers of deformed instantons in holographic baryon matter'',
J. High Energ. Phys. (2016) 2016, arXiv:1606.00675 [hep-ph].

\bibitem{Hoyos}
Carlos Hoyos, David Rodriguez Fernandez, Niko Jokela, Aleksi Vuorinen
``Holographic quark matter and neutron stars'', Phys. Rev. Lett. 117, 032501 (2016),
arXiv:1603.02943 [hep-ph].

\bibitem{FRE} 
Kazem Bitaghsir Fadafan, Jesus Cruz Rojas, Nick Evans
``Deconfined, Massive Quark Phase at High Density and Compact Stars: A Holographic Study'',
Phys. Rev. D 101, 126005 (2020), arXiv:1911.12705 [hep-ph]

\bibitem{Fadafan2}
Kazem Bitaghsir Fadafan, Jesus Cruz Rojas, Nick Evans, 
``Holographic quark matter with colour superconductivity and a stiff equation of state for compact stars'', Phys. Rev. D 103, 026012 (2021) [arXiv:2009.14079 [hep-ph]]. 

\bibitem{Nawa:2008uv}
K.~Nawa, H.~Suganuma and T.~Kojo,
Phys. Rev. D \textbf{79}, 026005 (2009)
doi:10.1103/PhysRevD.79.026005
[arXiv:0810.1005 [hep-th]].

\bibitem{Rho:2009ym}
M.~Rho, S.~J.~Sin and I.~Zahed,
Phys. Lett. B \textbf{689}, 23-27 (2010)
doi:10.1016/j.physletb.2010.01.077
[arXiv:0910.3774 [hep-th]].

\bibitem{Preis:2016fsp}
F.~Preis and A.~Schmitt,
JHEP \textbf{07}, 001 (2016)
doi:10.1007/JHEP07(2016)001
[arXiv:1606.00675 [hep-ph]].

\bibitem{Elliot-Ripley:2016uwb}
M.~Elliot-Ripley, P.~Sutcliffe and M.~Zamaklar,
JHEP \textbf{10}, 088 (2016)
doi:10.1007/JHEP10(2016)088
[arXiv:1607.04832 [hep-th]].

\bibitem{Ishii:2019gta}
T.~Ishii, M.~J\"arvinen and G.~Nijs,
JHEP \textbf{07}, 003 (2019)
doi:10.1007/JHEP07(2019)003
[arXiv:1903.06169 [hep-ph]].

\bibitem{Nakas:2020hyo}
T.~Nakas and K.~S.~Rigatos,
JHEP \textbf{12}, 157 (2020)
doi:10.1007/JHEP12(2020)157
[arXiv:2010.00025 [hep-th]].



\bibitem{Witt}
  E.~Witten,
  ``Anti-de Sitter space, thermal phase transition, and confinement in  gauge
  theories,''
  Adv.\ Theor.\ Math.\ Phys.\  {\bf 2} (1998) 505
  [arXiv:hep-th/9803131].

\bibitem{Ber}
 M. Berthelot,
 ``Sur quelques phenomenes de dilatation force des liquides,''
 Ann. Chim. Phys. {\bf 30} (1850) 332.

\bibitem{CLY}
Thomas D. Cohen, Scott Lawrence, Yukari Yamauchi,
``The thermodynamics of large-N QCD and the nature of metastable phases'',
Phys. Rev. C 102, 065206 (2020), arXiv:2006.14545 [hep-ph].


\bibitem{Demorest:2010bx}
P.~Demorest, T.~Pennucci, S.~Ransom, M.~Roberts and J.~Hessels,
Nature \textbf{467}, 1081-1083 (2010)
doi:10.1038/nature09466
[arXiv:1010.5788 [astro-ph.HE]].

\bibitem{Antoniadis:2013pzd}
J.~Antoniadis, P.~C.~C.~Freire, N.~Wex, T.~M.~Tauris, R.~S.~Lynch, M.~H.~van Kerkwijk, M.~Kramer, C.~Bassa, V.~S.~Dhillon and T.~Driebe, \textit{et al.}
Science \textbf{340}, 6131 (2013)
doi:10.1126/science.1233232
[arXiv:1304.6875 [astro-ph.HE]].

\bibitem{Annala:2017llu}
E.~Annala, T.~Gorda, A.~Kurkela and A.~Vuorinen,
Phys. Rev. Lett. \textbf{120}, no.17, 172703 (2018)
doi:10.1103/PhysRevLett.120.172703
[arXiv:1711.02644 [astro-ph.HE]].


\bibitem{De:2018uhw}
S.~De, D.~Finstad, J.~M.~Lattimer, D.~A.~Brown, E.~Berger and C.~M.~Biwer,
Phys. Rev. Lett. \textbf{121}, no.9, 091102 (2018)
[erratum: Phys. Rev. Lett. \textbf{121}, no.25, 259902 (2018)]
doi:10.1103/PhysRevLett.121.091102
[arXiv:1804.08583 [astro-ph.HE]].

\bibitem{Tews:2018iwm}
I.~Tews, J.~Margueron and S.~Reddy,
Phys. Rev. C \textbf{98}, no.4, 045804 (2018)
doi:10.1103/PhysRevC.98.045804
[arXiv:1804.02783 [nucl-th]].

\bibitem{Shibata:2017xdx}
M.~Shibata, S.~Fujibayashi, K.~Hotokezaka, K.~Kiuchi, K.~Kyutoku, Y.~Sekiguchi and M.~Tanaka,
Phys. Rev. D \textbf{96}, no.12, 123012 (2017)
doi:10.1103/PhysRevD.96.123012
[arXiv:1710.07579 [astro-ph.HE]].

\bibitem{Kojo:2020krb}
T.~Kojo,
AAPPS Bull. \textbf{31}, no.1, 11 (2021)
doi:10.1007/s43673-021-00011-6
[arXiv:2011.10940 [nucl-th]].

\bibitem{Prakash:1995uw}
M.~Prakash, J.~R.~Cooke and J.~M.~Lattimer,
Phys. Rev. D \textbf{52}, 661-665 (1995)
doi:10.1103/PhysRevD.52.661

\bibitem{Lattimer:2000nx}
J.~M.~Lattimer and M.~Prakash,
Astrophys. J. \textbf{550}, 426 (2001)
doi:10.1086/319702
[arXiv:astro-ph/0002232 [astro-ph]].

\bibitem{Ozel:2015fia}
F.~Ozel, D.~Psaltis, T.~Guver, G.~Baym, C.~Heinke and S.~Guillot,
Astrophys. J. \textbf{820}, no.1, 28 (2016)
doi:10.3847/0004-637X/820/1/28
[arXiv:1505.05155 [astro-ph.HE]].

\bibitem{Weise:2019mou}
W.~Weise,
JPS Conf. Proc. \textbf{26}, 011002 (2019)
doi:10.7566/JPSCP.26.011002
[arXiv:1905.03955 [nucl-th]].

\bibitem{GEK:2007}
J.~Erdmenger, K.~Ghoroku and I.~Kirsch,
``Holographic heavy-light mesons from non-Abelian DBI,''
J. High Energ. Phys. 09 (2007) 111, 
doi:10.1088/1126-6708/2007/09/111
[arXiv:0706.3978 [hep-th]].






 \end{thebibliography}
\end{document}